\documentclass[a4paper,12pt]{article}
\usepackage{a4wide, amsfonts, epsfig}

\newcommand{\sigmabf}{\mbox{\boldmath $\sigma$}}
\newcommand{\pibf}{\mbox{\boldmath $\pi$}}

\begin{document}

\title{\vskip -70pt
  \begin{flushright}
   {\normalsize TCDMATH 06-02, {\tt hep-th/0602227}}
  \end{flushright}
  \vskip 15pt {\bf The effect of pion mass on Skyrme configurations} \author{
{\large Conor Houghton and Shane Magee},\\ {\normalsize {\sl School of
Mathematics, Trinity College Dublin, Dublin 2, Ireland}}\\{\normalsize
{\sl Email: houghton@maths.tcd.ie, magees@maths.tcd.ie}}}
\date{{\large 26 September, 2006}}} \maketitle
\begin{abstract}
In the Skyrme model, atomic nuclei are identified with solitonic
configurations. If the pion mass is set to zero, these configurations
are spherical shells of energy with a fullerene-like appearance and
are well approximated by a simple rational map ansatz. Using simulated annealing, we have
 calculated minimum energy configurations for non-zero
pion mass and have found that they are less round and are less well
approximated by the rational map ansatz.
\end{abstract}

The Skyrme model \cite{S:1961} is a mathematically elegant effective
 theory which gives an approximate description of nucleons interacting
 at low energies. It is a pion field theory in which baryons appear as
 soliton configurations: the baryon number $B$ is given by the
 topological winding number of the soliton. The model has had some
 success in describing aspects of nuclear phenomenology such as the
 nucleon-nucleon potential \cite{JJP:1985} and the stability of $^4$He
 and $^7$Li \cite{BS:1997}. The model is non-renormalizable and, therefore,
 the approach usually taken is to treat it as a quantum mechanical
 model of nuclei and to quantize on a finite-dimensional space of
 classical configurations. For example, in \cite{ANW:1983}, the
 rotational space of minimum energy $B=1$ Skyrme configurations is
 quantized to give a reasonable description of the nucleon and the
 delta resonance.

The Skyrme field is an SU(2) field $U({\bf x})$ which is identified
with the normal pion fields $\pibf$ by
\begin{equation}
U=\exp{i\pibf \cdot \sigmabf},
\end{equation}
where $\sigmabf=(\sigma_1,\sigma_2,\sigma_3)$ are the usual Pauli
matrices. The energy functional is most conveniently written in terms
of the skew-Hermitian current $R_i=\partial_i U U^{\dagger}$ as
\begin{equation}\label{skmass}
E = \int \mbox{ d}^3 {\bf x} \
\left[ -\frac{1}{2} \mbox{ Tr} \left(R_i R_i \right)
    - \frac{1}{16}\mbox{ Tr} \left( [ R_i, R_j][R_i, R_j]\right) 
-m^2 \mbox{ Tr} \left(U -1 \right) \right].
\end{equation}
Here, the coupling constants appearing in front of the first two terms
have been scaled to one by a choice of the length and energy
scales. The Skyrme model is a modified sigma model and the first term
in the energy is the usual sigma-model energy. The second term was
added by Skyrme \cite{S:1961} and is needed to prevent Derrick scaling \cite{D:1964} of the minimum energy solutions.  
The chiral symmetry-breaking third term was introduced in
\cite{AN:1984}; writing the Lagrangian in terms of the pion fields, $\pibf$, shows that $m$ corresponds to the pion mass.

Most classical investigations of the Skyrme model have ignored the
effect of the pion mass. However, in the model the classical
counterpart of nucleons are rotating configurations and stable
rotating Skyrme configuration must have a non-zero pion mass
\cite{RSWW:1986}. In fact, the faster the rotation, the larger the
pion mass required. In order to produce stable Skyrme configurations
that model the nucleon and delta resonance, the pion mass must be set
to over twice its experimental value \cite{BKS:2005}.

Skyrme configurations with zero pion mass have a remarkable
fullerene-like appearance: their energy density is concentrated along
the edges and vertices of a polyhedral skeleton. Furthermore, they are
well approximated by the rational map ansatz \cite{HMS:1998} which restricts Skyrme fields to the form:
\begin{equation}\label{fldansatz}
U=\exp(-if(r){\mathbf n}(\theta,\phi) \cdot \sigmabf),
\end{equation}
where $f(r)$ is a radial profile function and ${\mathbf
n}(\theta,\phi)$ is a unit vector with a specific holomorphic
form, which is described in \cite{HMS:1998}. The ansatz fields have a very
distinctive structure; the iso-surfaces of trace $U$, for example, are
spheres around the origin.

If this field ansatz is substituted into the Skyrme energy functional,
the minimization problem splits into two parts: one for the angular
behavior, allowing the angular map ${\bf n}(\theta,\phi)$ to be
calculated, and a second for the radial behavior, allowing $f(r)$ to
be calculated. Only the second of these is affected by the pion mass.
If the ansatz is used to approximate minimum energy Skyrme
configurations for non-zero pion mass the angular behavior is
identical to the angular behavior for zero pion mass. The only change
is in the profile function: for non-zero pion mass it has a more rapid
fall-off resulting in a smaller, tighter, configuration. However, in
\cite{BS:2005} it is argued that the rational map ansatz cannot
produce good approximations to the minimum energy solutions for large
pion mass and large baryon numbers. In fact, it is shown here that the rational map ansatz gives a less accurate
approximation for non-zero pion mass and the iso-surfaces of trace $U$ are not spherical.

Here, the minimum energy Skyrme configurations are calculated
numerically for a range of values of the pion mass $m$. This is performed
using simulated annealing: an established method of generating minimum
energy Skyrme configurations \cite{HSW:2000,LM:2005}. However,
simulated annealing can be computationally expensive and this limits
the possible baryon numbers that can be considered. High pion mass is
an advantage here because Skyrme configurations with high pion mass
are much smaller than their counterparts with zero pion mass. For
example, using a value of $m=3.8$, configurations up to $B=9$ can be
annealed on a 100$^3$ lattice with lattice spacing of 0.06 Skyrme units
without significant edge effects. The pion mass value of $m=3.8$ is
approximately eight times the experimental value and three times that
used in \cite{BKS:2005} to ensure stable rotating solutions. It is
high enough to give a clear qualitative demonstration of pion mass
effects; the best value of the pion mass to use in a putative Skyrme
model of nuclear physics is an open question. 

One disadvantage of using a higher pion mass is that tighter Skyrme
configurations have very high field derivatives. These are less well
approximated by the lattice Taylor expansion used in the lattice
simulation. For example, a good test of numerical accuracy is to
calculate the baryon number which should be an integer. If we adopted
the 0.12 lattice spacing used in \cite{HSW:2000} and \cite{LM:2005},
the error in baryon number is approximately 8\% for the $m=3.8$
solution compared with 2\% for the $m=0$ case. A reduction in lattice
spacing to 0.06 reduces the error to less than 2\%, and this is the
lattice spacing we have used to produce the results presented here.

\begin{figure}
\begin{center}
\includegraphics[scale=0.65]{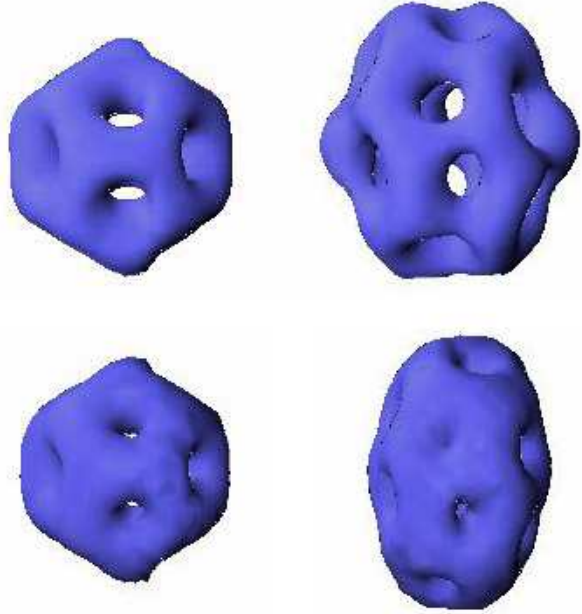}
\end{center}
\caption{Iso-surface plots at baryon density ${\mathcal B}=.017$ and pion mass $m$=3.8 for baryon numbers seven and nine. The top row are the annealed solutions obtained from rational map initial conditions while the bottom row are their corresponding rational map ansatz approximations.}
\label{b7and9}
\end{figure}

\begin{figure}
\begin{center}
\includegraphics[scale=0.8]{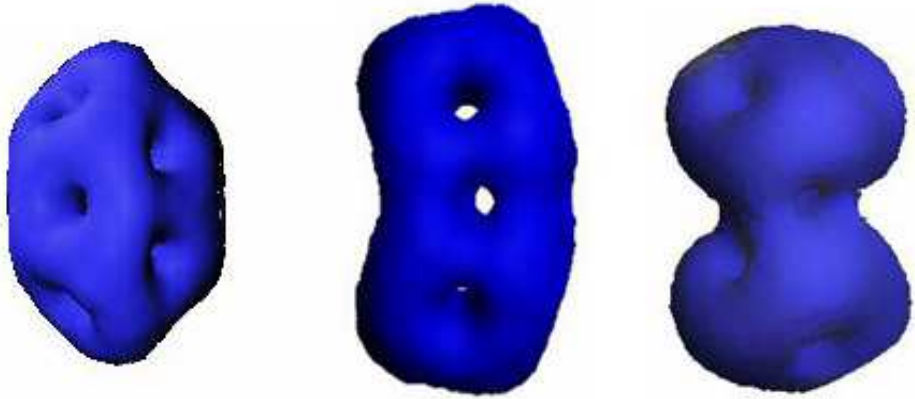}
\end{center}
\caption{Three local $B=8$ minima for pion mass $m=3.8$, obtained from
three different initial configurations, from left, a $B=8$ rational
map, a $B=4$ rational map wound twice around the target sphere, and
eight separated $B=1$ Skyrmions. These configurations have energies,
from left, of 1.9, 1.91 and 1.917 in $12 \pi^2 B$ Skyrme units}
\label{B8}
\end{figure}

Using simulated annealing, we have calculated minimum energy
configurations for the Skyrme model with pion mass $m=3.8$. Baryon
density iso-surfaces are shown in Fig. \ref{b7and9} for $B=7$ and 9,
and Fig. \ref{B8} for $B=8$. By repeating the minimization for a range
of pion masses, we have verified that the effect seen here varies in
quantity, but not in its general form, as the pion mass changes.

\begin{figure}
\begin{center}
\includegraphics[scale=0.38]{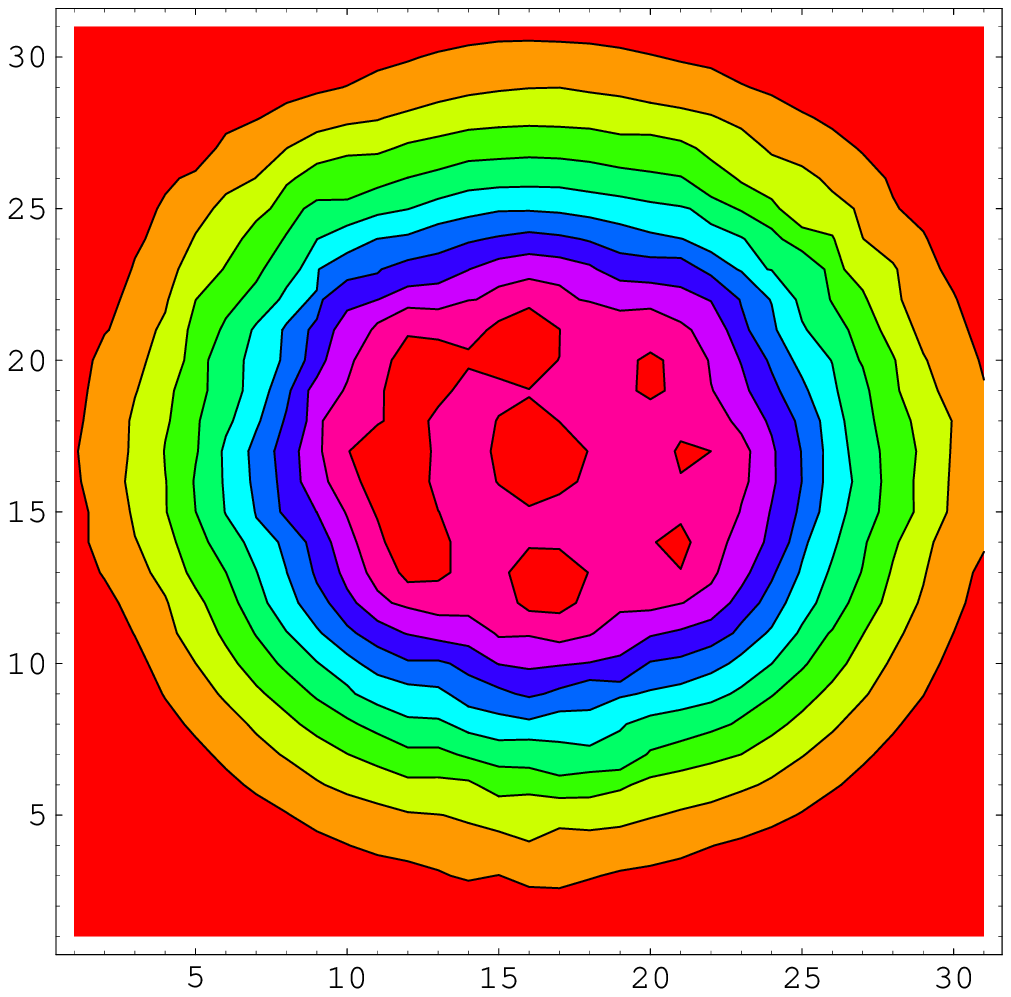}
\includegraphics[scale=0.38]{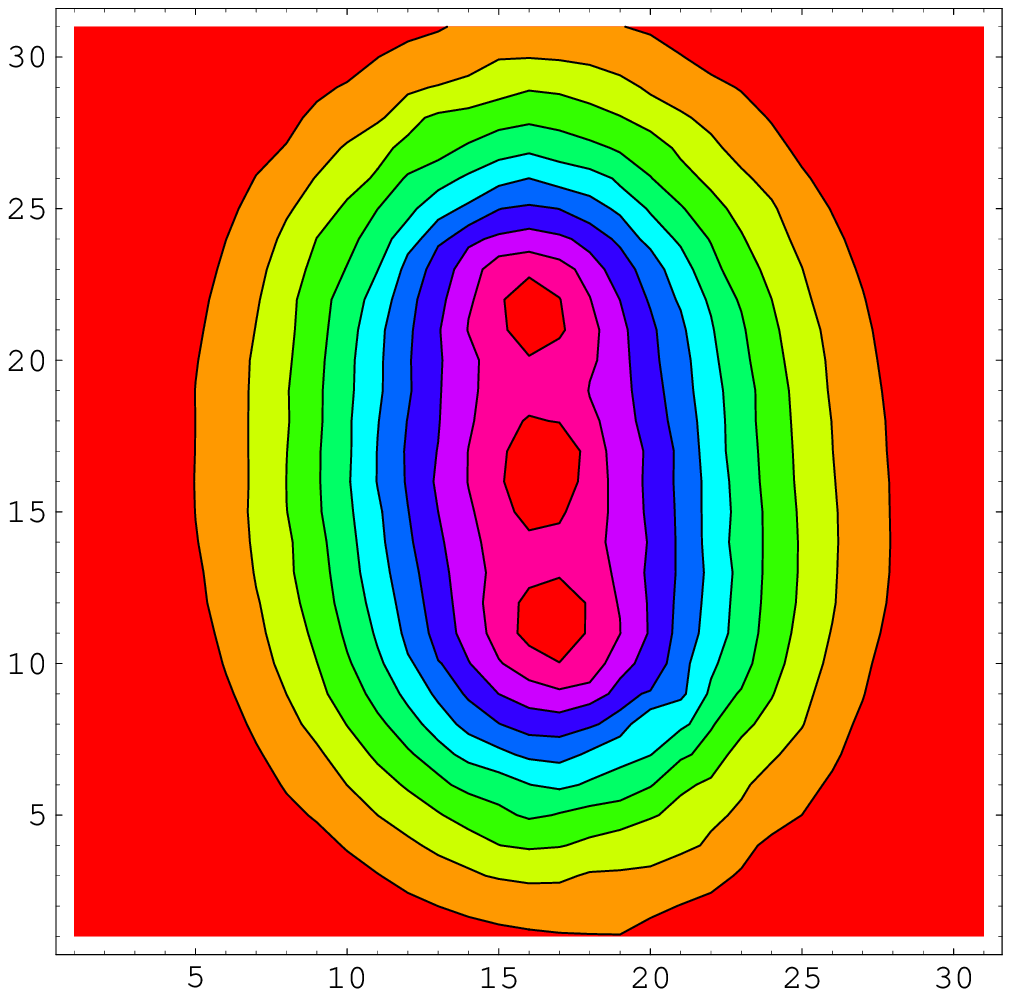}
\includegraphics[scale=0.38]{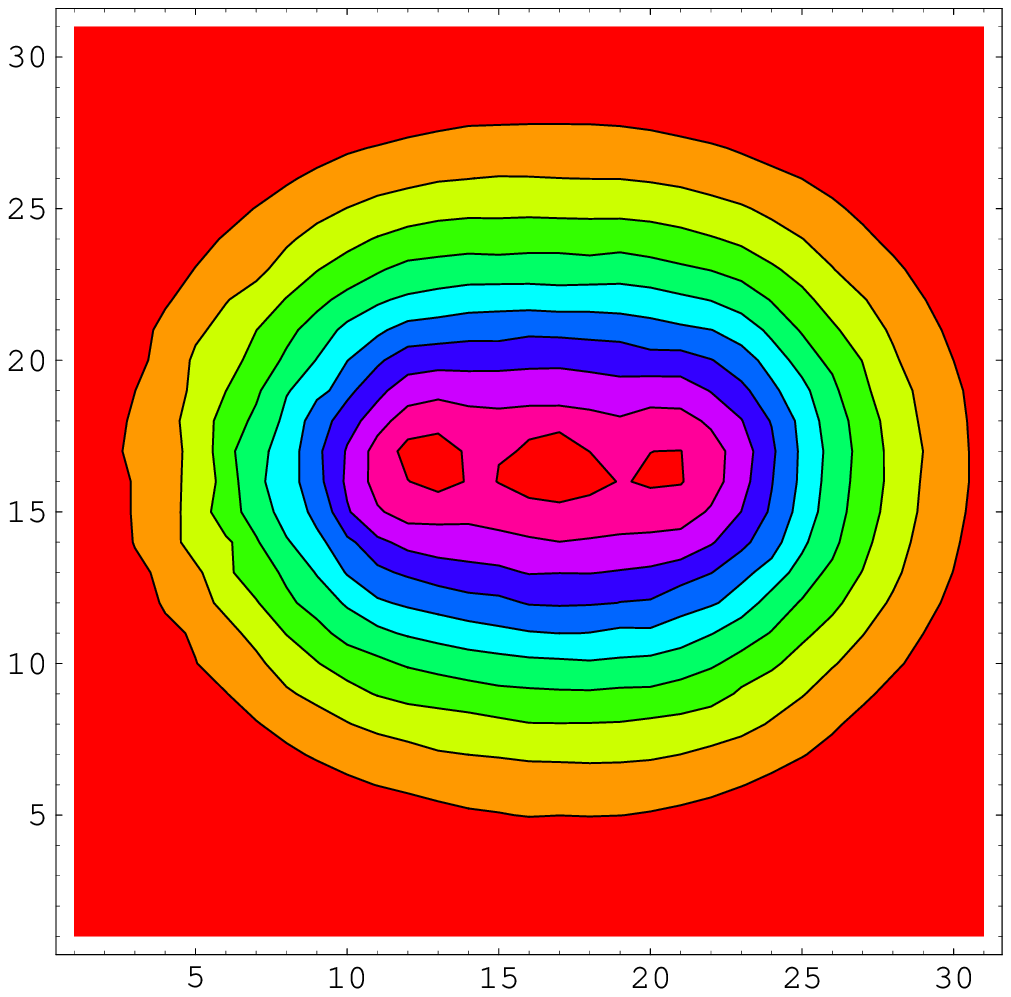}
\includegraphics[scale=0.38]{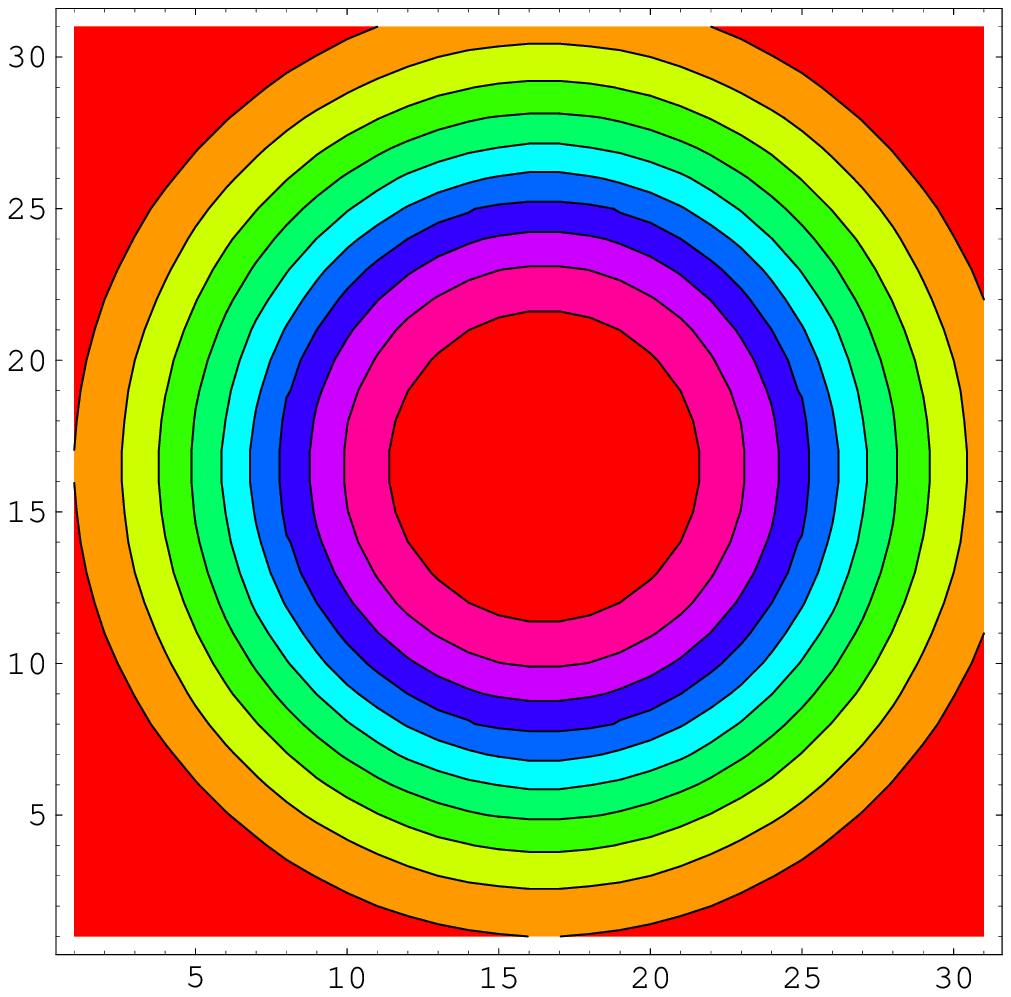}
\end{center}
\caption{Contour plot of trace$\,U$ of an annealed
$B=9$ configuration with $m=3.8$ and a lattice spacing of .06 Skyrme
units. The first three plots are cross-sections through the origin in
the $x$, $y$ and $z$ directions, while the plot on the very right is
the spherical cross-section for the rational map configuration. }
\label{B9pr}
\end{figure}

The $B=7$ configuration has the same shape for zero and non-zero pion
mass.  With this exception, the distinguishing feature of the
solutions found so far is that the configuration is flatter. This can
be understood physically: the region inside the fullerene-like shell
has a Skyrme field close to the anti-vacuum value, $U\approx-{\bf
1}$. Such field values have a high pion-mass density and so the
interior of the shell contributes to the energy if the pion mass is
non-zero. A flatter configuration has a small interior volume for a
particular surface area. Figure \ref{B9pr} shows contours of
trace$\,U$ for planar cross-sections through the $B=9$
configuration. Unlike the zero-pion mass configuration, the contours
are not circular. The rational map ansatz relies on a spherical polar
decomposition of ${\bf R}^3$. It is possible that another
decomposition of space is appropriate for Skyrmions with non-zero pion
mass. For example, it is already known that Skyrmions with non-zero
pion mass are related to Skyrmions with zero-pion mass in hyperbolic
space \cite{AS:2005}.

For baryon numbers seven and nine, the point symmetry group of the
minimum energy solution remains unchanged; the situation is less
clear-cut for $B=8$. We originally obtained a minimum energy result
for $B=8$ that had the same point group as its $m=0$
counterpart. This, however, seems to belong to a family of minima with
similar energy. In \cite{BS:2006}, using a different numerical scheme,
a different minimum energy solution was obtained.  Furthermore, the addition
of the pion mass has the effect of considerably raising the energy
barrier between minima and reducing the portion of Skyrme
configuration space that can be searched by the algorithm in a
reasonable time.

This is a common hurdle in simulated annealing applications and is
generally dealt with using what is called the multi-start strategy,
performing many runs, each using a different initial conditions.
Although, in principle, simulated annealing will find a global minimum
with a sufficiently slow cooling schedule, in practice, even a very
slow schedule may not resolve local minima with proximate energy
values. In our implementation, the initial condition is important when
minimizing $B=8$ configurations. We have therefore tried initializing
the algorithm with many different starting configurations. Three
minima have been found which our scheme cannot resolve within
numerical accuracy: these are shown in Fig. \ref{B8}. In addition to a
solution with the same point group as the $m=0$ equivalent, there are
also other solutions which resemble the alpha-clusters discussed in
\cite{BMS:2006}.

One outstanding problem with attempts to model nuclei as Skyrmions is
that for many values of $B$ the semi-classical quantization of the
classical minimum does not give the correct quantum spin and iso-spin
numbers \cite{I:2000,K:2003}. This is a result of the large amount of
symmetry possessed by the classical minimum. It might have been hoped
that the minimum energy solutions with non-zero pion mass would be
less symmetric. This does, indeed, seem to be the case; however, for
$B=7$, where the discrepancy between computed and experimental quantum
numbers is most glaring, there is no difference in shape between the
zero and non-zero pion mass solutions. It has been suggested that a
quasi-classical quantization which also takes greater account of the effect
of rotation might yield the correct quantum numbers. A prescription
for such a quantization is given in \cite{HM:2005} and numerical work
is continuing in this direction.

\section*{Acknowledgements}
SM acknowledges receipt of funding under the Programme for Research in Third Level Institutions (PRTLI), administered by the HEA; CJH and SM acknowledge receipt of a Trinity College Dublin start-up grant. We would also like to thank the Trinity Centre for High Performance Computing and the Irish Centre for High End Computing for the use of their computing facilities. We would like to thank Richard Battye and Paul Sutcliffe for sharing early results of their paper \cite{BS:2006} when our own paper was in draft form.

\end{document}